\documentclass[useAMS,usenatbib,preprint2,flushrt]{emulateapj}
\usepackage{epsfig}
\usepackage{amsmath}
\usepackage{amssymb}
\usepackage{mathrsfs}
\usepackage{float}
\usepackage{graphicx}

\newcommand{\beq}{\begin{equation}}
\newcommand{\eeq}{\end{equation}}
\newcommand{\msun}{$M_{\odot}\hspace{1mm}$}

\newcommand{\Awseven}{1.40^{+0.31}_{-0.31}}
\newcommand{\Awsevenbright}{1.77^{+0.82}_{-0.80}}
\newcommand{\Awsevenfaint}{1.24^{+0.43}_{-0.43}}
\newcommand{\roseven}{6.7^{+0.9}_{-1.0}}
\newcommand{\rosevenbright}{7.8^{+2.1}_{-2.5}}
\newcommand{\rosevenfaint}{6.2^{+1.3}_{-1.4}}
\newcommand{\biasseven}{8.6^{+0.9}_{-1.0}}
\newcommand{\biassevenbright}{9.7^{+2.0}_{-2.5}}
\newcommand{\biassevenfaint}{8.1^{+1.3}_{-1.6}}
\newcommand{\halomassseven}{11.1^{+0.2}_{-0.3}}

\newcommand{\AwsevenBouwensOnly}{1.5^{+0.4}_{-0.4}}
\newcommand{\rosevenBouwensOnly}{7.0^{+1.1}_{-1.2}}
\newcommand{\biassevenBouwensOnly}{8.9^{+1.1}_{-1.3}}
\newcommand{\Awsixpointeight}{1.6^{+0.4}_{-0.4}}
\newcommand{\rosixpointeight}{5.9^{+0.9}_{-1.0}}
\newcommand{\biassixpointeight}{7.4^{+0.9}_{-1.0}}
\newcommand{\Awsevenpointeight}{1.7^{+0.9}_{-0.8}}
\newcommand{\rosevenpointeight}{6.1^{+1.8}_{-2.1}}
\newcommand{\biassevenpointeight}{8.6^{+2.0}_{-2.5}}

\newcommand{\Awsix}{0.93^{+0.34}_{-0.34}}
\newcommand{\Awsixbright}{1.79^{+0.77}_{-0.76}}
\newcommand{\Awsixfaint}{0.34^{+0.35}_{-0.23}}
\newcommand{\rosix}{3.6^{+0.8}_{-0.9}}
\newcommand{\rosixbright}{5.5^{+1.4}_{-1.6}}
\newcommand{\rosixfaint}{1.9^{+1.1}_{-1.0}}
\newcommand{\biassix}{4.4^{+0.8}_{-0.9}}
\newcommand{\biassixbright}{6.2^{+1.2}_{-1.5}}
\newcommand{\biassixfaint}{2.7^{+1.2}_{-1.2}}

\newcommand{\Awfive}{0.60^{+0.13}_{-0.13}}
\newcommand{\Awfivebright}{0.68^{+0.27}_{-0.26}}
\newcommand{\Awfivefaint}{0.54^{+0.22}_{-0.22}}
\newcommand{\rofive}{3.4^{+0.4}_{-0.5}}
\newcommand{\rofivebright}{3.7^{+0.9}_{-1.0}}
\newcommand{\rofivefaint}{3.2^{+0.8}_{-0.9}}
\newcommand{\biasfive}{3.6^{+0.4}_{-0.4}}
\newcommand{\biasfivebright}{3.8^{+0.7}_{-0.8}}
\newcommand{\biasfivefaint}{3.4^{+0.6}_{-0.8}}

\newcommand{\Awfour}{0.44^{+0.05}_{-0.05}}
\newcommand{\Awfourbright}{0.80^{+0.10}_{-0.10}}
\newcommand{\Awfourfaint}{0.24^{+0.08}_{-0.08}}
\newcommand{\rofour}{3.5^{+0.2}_{-0.2}}
\newcommand{\rofourbright}{5.1^{+0.4}_{-0.4}}
\newcommand{\rofourfaint}{2.4^{+0.5}_{-0.5}}
\newcommand{\biasfour}{3.0^{+0.2}_{-0.2}}
\newcommand{\biasfourbright}{4.0^{+0.3}_{-0.3}}
\newcommand{\biasfourfaint}{2.2^{+0.3}_{-0.4}}

\shorttitle{Galaxy clustering at $\lowercase{z}  \sim 7.2$}
\shortauthors{Barone-Nugent et al.}

\begin{document}

\title{Measurement of galaxy clustering at $\lowercase{z} \sim 7.2$ and the evolution of galaxy bias from $3.8 < \lowercase{z} < 8$ in the XDF, GOODS-S and GOODS-N}
   
\author{R. L. Barone-Nugent\altaffilmark{1}, M. Trenti\altaffilmark{2}, J. S. B. Wyithe\altaffilmark{1}, R. J. Bouwens\altaffilmark{3}, P. A. Oesch\altaffilmark{4,5}, G. D. Illingworth\altaffilmark{4}, C. M. Carollo\altaffilmark{6}, J. Su\altaffilmark{7}, M. Stiavelli\altaffilmark{8}, I. Labbe\altaffilmark{3}, P. G. van Dokkum\altaffilmark{9}}

\altaffiltext{1}{School of Physics, University of Melbourne, Parkville, Victoria, Australia}
\altaffiltext{2}{Kavli Institute for Cosmology and Institute of Astronomy, University of Cambridge, Cambridge, United Kingdom}
\altaffiltext{3}{Leiden Observatory, Leiden University, NL-2300 RA Leiden, The Netherlands}
\altaffiltext{4}{UCO/Lick Observatory, University of California, Santa Cruz, CA 95064, USA}
\altaffiltext{5}{Yale Center for Astronomy and Astrophysics, Yale University, PO Box 208121, New Haven, CT 06520, USA}
\altaffiltext{6}{Institute for Astronomy, ETH Zurich, 8092 Zurich, Switzerland}
\altaffiltext{7}{Department of Radiology, Mayo Clinic, Rochester, MN 55905, USA}
\altaffiltext{8}{Space Telescope Science Institute, Baltimore, MD 21218, USA}
\altaffiltext{9}{Department of Astronomy, Yale University, New Haven, CT 06520, USA}

\begin{abstract}
 
Lyman-Break Galaxy (LBG) samples observed during reionization ($z\gtrsim6$) with Hubble Space Telescope's WideFieldCamera3 are reaching sizes sufficient to characterize their clustering properties. Using a combined catalog from the Hubble eXtreme Deep Field and CANDELS surveys, containing $N=743$ LBG candidates at $z\geq6.5$ at a mean redshift of $\overline{z}=7.2$, we detect a clear clustering signal in the angular correlation function (ACF) at $\gtrsim4\sigma$, corresponding to a real-space correlation length $r_0=\roseven h^{-1}$cMpc. The derived galaxy bias $b=\biasseven$ is that of dark-matter halos of $M=10^{\halomassseven}$\msun at $z=7.2$, and highlights that galaxies below the current detection limit ($M_{AB}\sim-17.7$) are expected in lower-mass halos ($M\sim10^8-10^{10.5}$\msun). We compute the ACF of LBGs at $z\sim3.8-z\sim5.9$ in the same surveys. A trend of increasing bias is found from $\overline{z}=3.8$ ($b\sim3.0$) to $\overline{z}=7.2$ ($b\sim8.6$), broadly consistent with galaxies at fixed luminosity being hosted in dark-matter halos of similar mass at $4\lesssim z\lesssim6$, followed by a slight rise in halo masses at $z\gtrsim7$ ($\sim2\sigma$ confidence). Separating the data at the median luminosity of the $\overline{z}=7.2$ sample ($M_{UV}=-19.4$) shows higher clustering at $\overline{z}=5.9$ for bright galaxies ($r_0=\rosixbright h^{-1}$cMpc, $b=\biassixbright$) compared to faint galaxies ($r_0=\rosixfaint h^{-1}$cMpc, $b=\biassixfaint$) implying a constant mass-to-light ratio $\frac{\textnormal{dlogM}}{\textnormal{dlogL}}\sim1.2^{+1.8}_{-0.8}$. A similar trend is present in the $\overline{z}=7.2$ sample with larger uncertainty. Finally, our bias measurements allow us to investigate the fraction of dark-matter halos hosting UV-bright galaxies (the duty-cycle, $\epsilon_{\textnormal{DC}}$). At $\overline{z}=7.2$ values near unity are preferred, which may be explained by the shortened halo assembly time at high-redshift.

\end{abstract}
%

\section{Introduction}

The large-scale structure of the Universe and the physics of galaxy formation during the epoch of Reionization are now accessible to observational tests thanks to the Hubble Space Telescope WideFieldCamera3 (WFC3). The new WFC3 transformed the field by providing the first large samples of galaxies seen in the first $500-700$ Myr after the Big Bang (redshift $z\sim7-10$), identified using the Lyman-break, or dropout, technique \citep{steidel1996}, from a variety of surveys \citep{bouwens2014,windhorst2011,trenti2011, mclure2012, bradley2012, oesch2012}. Such $z\geq6.5$ observations show that there is a decreasing number density of galaxies as the epoch of reionization is approached, albeit with growing evidence of a steepening of the galaxy luminosity function (LF) faint end \citep{bouwens2014}. By assuming that brighter galaxies are hosted in more massive dark-matter halos, the evolution of the galaxy LF can be linked to the underlying dark-matter halo mass function. Empirical models \citep{trenti2010,tacchella2013}, analytic models \citep{wyithe2013suppressed} and cosmological simulations \citep{jaacks2012, lacey2011} are all successful at reproducing the star-formation rate and LF, but there is little validation of predictions beyond galaxy number counts.

Studying the clustering properties of galaxies, through their Angular Correlation Function (ACF), is an avenue to probe the connection between the observed light and host dark-matter halos. At lower redshifts, extensive studies of clustering have been carried out, leading to measurements of the galaxy bias (the extra clustering of galaxies/halos compared to that of dark-matter), and host dark-matter halo mass \citep{adelberger1998, arnouts1999, giavalisco2001, ouchi2004, lee2006, overzier2006}. 

\citet{lee2006} investigated the observed distribution of the number of galaxies in a dark-matter halo at higher redshifts, known as the halo-occupation distribution, finding evidence for multiple occupation and sub-halo clustering, extending similar studies at lower redshift (e.g., \citealt{berlind2002}). 

Furthermore, clustering enables characterization of the expected evolutionary history of the observed galaxy population. For example, \citet{adelberger2005} compared their analysis of clustering of $z\lesssim3.5$ LBGs with analyses of clustering at $z=0$ to infer that their LBG sample represents the progenitors of present-day ellipticals. We can now take advantage of the large sample size and characterize the clustering properties of high-redshift Lyman-Break Galaxies (LBGs).

In this paper, we measure LBG clustering at $\overline{z}=7.2$ for the first time. We investigate qualitative trends of average bias and clustering strength with luminosity, compare our results with those at lower redshift, constrain the fraction of dark-matter halos hosting LBGs (the duty cycle), and test high-$z$ theoretical modeling of halo occupation \citep{trenti2010, wyithe2013}. This paper is organized as follows: Section \ref{section:data} describes the LBG sample. Section \ref{section:tpcf} presents our measurements of the ACF, and Section \ref{section:results} our results. Section \ref{section:conclusion} summarizes and concludes. We use the latest Planck cosmology ($\Omega_{M}$,$\Omega_{\lambda}$,$h$,$\sigma_{8}$) = ($0.315$, $0.685$, $0.673$, $0.828$) \citep{collaboration2013}. Magnitudes are in the AB system \citep{oke1974}.
%

\section{Data}
\label{section:data}
Our analysis is based on a combined sample of $\overline{z}=7.2$ candidate galaxies, photometrically selected as $z$ and $Y$-band dropouts from the \citet{bouwens2014} [B+14] and \citet{mclure2012} analysis of Hubble's deep and ultradeep area on the XDF/UDF fields \citep{illingworth2013} and the CANDELS data over GOODS-North and South (GN and GS, respectively; \citealt{grogin2011}). The observations span from the $4.7$ arcmin$^2$ area of the XDF, which reaches $m_{H_{160}}=29.8$ ($5\sigma$), to the $\sim280$ arcmin$^2$ of CANDELS at $m_{H_{160}}\sim27.7-28.5$. The B+14 catalogs include $N=670$ objects at $z>6.5$. The \citet{mclure2012} catalogs list $N=197$ objects within XDF/GOODS-South. Galaxies in each catalog are selected using different methods; B+14 use a two-color selection to detect the Lyman-break, while \citet{mclure2012} use multi-band spectral energy distribution fitting. A combined catalog is constructed by removing double-counted objects, and by counting the 31 pairs with separation $d<0\farcs3$ ($5$ pixels, $\sim9$kpc at $z=7.2$) as single sources, with their fluxes combined since such small separations are comparable to the HST point spread function in the $H$-band. This yields a total sample of $N=743$ objects, with $\sim80\%$ overlap for objects one magnitude above the survey detection limit, and $\sim40\%$ overlap for less robust candidates.
To ensure that merging the two catalogs does not bias our ACF measurement, we carry out a control analysis of the ACF using only the B+14 catalog (Section \ref{section:results}).

We complement our new study at $\overline{z}=7.2$, with a re-analysis of clustering at lower redshift, using the latest $B$, $V$ and $i$-dropout ($z\sim3.8-6.0$ catalogs by B+14. Table~\ref{table:fields} lists our samples, which we split between ``bright'' and ``faint'' subsamples at the median magnitude of the $\overline{z}=7.2$ sample, $M_{UV}=-19.4$, to study the redshift evolution of the bias at constant UV luminosity. A comprehensive description of the data, including number counts, luminosity functions, and redshift distributions of the samples is available in B+14.

\begin{table*}
\begin{center}
\begin{tabular}{c | c | c | c | c}
\hline
\hline
Field & $\overline{z}=3.8$ & $\overline{z}=4.9$ & $\overline{z}=5.9$ & $\overline{z}=7.2$\\
\hline
XDF			&\textbf{385}(\textbf{57})	&\textbf{157}(\textbf{19})		&\textbf{104}(\textbf{13})		&\textbf{149}(\textbf{11})\\
HUDF091		&---						&\textbf{93}(\textbf{24})		&\textbf{38}(\textbf{9})		&\textbf{52}(\textbf{9})\\
HUDF092		&\textbf{147}(\textbf{50})	&\textbf{83}(\textbf{17})		&\textbf{36}(\textbf{10})		&\textbf{54}(\textbf{8})\\
GS-Deep		&\textbf{1621}(\textbf{756})&\textbf{537}(\textbf{251})		&\textbf{203}(\textbf{102})		&\textbf{134}(\textbf{76})\\
GS-ERS		&\textbf{757}(\textbf{445})	&\textbf{209}(\textbf{150})		&\textbf{62}(\textbf{49})		&\textbf{64}(\textbf{52})\\
GS-Wide		&\textbf{422}(\textbf{304})	&\textbf{122}(\textbf{98})		&\textbf{41}(\textbf{36})		&13(11)\\
GN-Deep		&\textbf{1721}(\textbf{770})&\textbf{745}(\textbf{352})		&\textbf{197}(\textbf{108})		&\textbf{220}(\textbf{151})\\
GN-Wide 1	&\textbf{398}(\textbf{302})	&\textbf{138}(\textbf{116})		&25(24)							&33(33)\\
GN-Wide 2	&\textbf{485}(\textbf{346})	&\textbf{173}(\textbf{130})		&\textbf{51}(\textbf{44})		&24(24)\\
\hline
TOTAL		& 5936(3030)	&2257(1157)	&757(395)	&743(375)\\
\hline
\end{tabular}
\caption{Number of LBGs (bright LBGs) in each field at each redshift. The data are divided into bright and faint subsamples based on the median luminosity of the $\overline{z}=7.2$) sample $M_{UV}=-19.4$, which is within $\sim0.1$ mag of the median of samples at lower-$z$. Only entries in \textbf{bold} are used in the ACF analysis, since we discard fields where mean galaxy separation exceeds $>100\farcs0$.}
\label{table:fields}
\end{center}
\end{table*}
%

\begin{table*}
\begin{center}
\begin{tabular}{c | c c c |c c c |c c c}
\hline
\hline
$\overline{z}$ & $A_{w,total}$ &  $r_{0, total}$ & $b_{total}$ & $A_{w,bright}$ & $r_{0, bright}$ & $b_{bright}$ & $A_{w,faint}$ & $r_{0, faint}$ & $b_{faint}$\\
\hline
$3.8$ 	&  $\Awfour$ & $\rofour$   	& $\biasfour$  	& $\Awfourbright$	& $\rofourbright$	& $\biasfourbright$ & $\Awfourfaint$	& $\rofourfaint$	& $\biasfourfaint$\\
$4.9$ 	&  $\Awfive$ & $\rofive$   	& $\biasfive$  	& $\Awfivebright$   & $\rofivebright$	& $\biasfivebright$ & $\Awfivefaint$   	& $\rofivefaint$	& $\biasfivefaint$\\
$5.9$ 	&  $\Awsix$  & $\rosix$   	& $\biassix$ 	& $\Awsixbright$  	& $\rosixbright$	& $\biassixbright$ & $\Awsixfaint$  	& $\rosixfaint$		& $\biassixfaint$\\
$7.2$ 	&  $\Awseven$ & $\roseven$   & $\biasseven$	& $\Awsevenbright$  & $\rosevenbright$	& $\biassevenbright$ & $\Awsevenfaint$  & $\rosevenfaint$	& $\biassevenfaint$\\
\hline
\end{tabular}
\caption{The free ACF parameter, $A_{w}$ ($\beta=0.6$ is fixed), real-space correlation length and galaxy bias for the total, bright and faint samples at each redshift. Units of $r_{0}$ are $h^{-1}$Mpc.}
\label{table:values}
\end{center}
\end{table*}

\section{Estimating the angular correlation function and bias}
\label{section:tpcf}
The ACF measures the excess probability of finding two galaxies at an angular separation $\theta$, over a random Poisson point process. We use the ACF estimator, $w(\theta)=(DD(\theta)-2DR(\theta)+RR(\theta))/RR(\theta)$ \citep{landy1993}, where $DD(\theta)$, $DR(\theta)$ and $RR(\theta)$ are the number of galaxy-galaxy pairs, galaxy-random point pairs and random-random point pairs (respectively) counted at a separation of $\theta\pm\delta\theta$. We produced random point catalogs using a spatial Poisson process, accounting for field geometry and crowding effects using segmentation maps produced by $\tt{SExtractor}$ \citep{bertin1996}.

We use linear binning with width $12\farcs5$ to construct the ACF for the full sample. For subsamples at $\overline{z}=5.9$ and $\overline{z}=7.2$ we increase bin-width to $25\farcs0$ to account for the reduced number of pairs. We assume a power law parameterization of the ACF, $w(\theta)=A_{w}\theta^{-\beta}$. Following previous investigations \citep{lee2006,overzier2006}, we fix $\beta=0.6$, but we quantify in Section \ref{section:results} the systematic uncertainty associated with this choice. Because of finite survey area, the observed ACF is underestimated by a constant known as the integral constraint (IC):
\beq
w_{\textnormal{true}}(\theta)=w_{\textnormal{obs}}(\theta)+\textnormal{IC}.
\eeq
IC can be estimated from:
\begin{align}
\textnormal{IC}&=\frac{1}{\Omega^{2}}\int_{1}\int_{2}w_{\textnormal{true}}(\theta)d\Omega_{1}d\Omega_{2}\\
&=\frac{\sum_{i}RR(\theta_{i}) w_{\textnormal{true}}(\theta_{i}) }{\sum_{i}RR(\theta_{i}) }=\frac{\sum_{i}RR(\theta_{i})A_{w}\theta_{i}^{-\beta}}{\Sigma_{i}RR(\theta_{i})},
\end{align}
where $w_{\textnormal{true}}(\theta)$ is the intrinsic ACF and $w_{\textnormal{obs}}(\theta)$ is the measurement within the survey area, $\Omega$. For small area fields such as the XDF, the integral constraint is $0.96A_{w}$, while in larger fields (GOODS-S Deep) the IC starts becoming a second-order correction ($IC=0.04A_{w}$). The only uncertainty in the IC derives from assuming fixed $\beta$. $A_{w}$ is obtained from the observed ACF in the fields (each with their own IC) by maximizing the likelihood given by: 
\begin{equation}
\mathscr{L}=\prod_{i=\textrm{fields}}\frac{1}{\sigma(\theta)_{obs,i}\sqrt{2\pi}} e^{-\frac{1}{2}\Big(\frac{w(\theta)_{obs,i}-A_{w}(\theta^{-\beta}-\frac{\textrm{IC}_{i}}{A_{w}})}{\sigma(\theta)_{obs,i}}\Big)^{2}}
\label{eq:fit}
\end{equation}
{where $w(\theta)_{obs,i}$, IC$_{i}$ and $\sigma(\theta)_{i}$ are the ACF measurements, integral constraints and uncertainties in field $i$, respectively.
After $A_{w}$ is determined, IC values can be added to the ACF measurements in each field, and combined to determine the intrinsic ACF shown in Figure \ref{figure:tpcf-6.5+}. Errors in the ACF are estimated using bootstrap resampling \citep{ling1986}, without including systematic sample (``cosmic'') variance uncertainty, which might be comparable to the random error (e.g., \citealt{trenti08}, and see Section~\ref{section:results} for field-to-field variations in our analysis).

We approximate the real-space correlation function with a power law $\xi(r)=(r/r_{0})^{-\gamma}$, where the coefficients are related to the ACF coefficients by the Limber transform \citep{peebles1980, adelberger2005},
\begin{align}
&\beta=\gamma-1,\\
&A_{w}=\frac{r_{0}^{\gamma} B[1/2,(\gamma-1)]\int_{0}^{\infty}dzN(z)^{2}f^{1-\gamma}g(z)^{-1}}{\big[\int_{0}^{\infty} N(z)\big]^{-2}}.
\end{align}
Here $f\equiv(1+z)D_{A}(\theta)$ where $D_{A}(\theta)$ is the angular diameter distance and $N(z)$ is the redshift distribution of the dropouts, $B$ is the beta function and $g(z)\equiv c/H(z)$. We assume the clustering evolution to be fixed in comoving coordinates over each redshift window. $N(z)$ is taken from B+14 (see their Figure~1 for quantitative details), with mean values reported in Table~\ref{table:values}, and typical standard deviations $\approx0.4$. From the real-space correlation function, we define the galaxy bias $b$ as the ratio of the galaxy variance at $8h^{-1}$cMpc (comoving megaparsecs), $\sigma_{8,g}$, to the linear matter fluctuation at $8h^{-1}$cMpc, $\sigma_{8}$:
\begin{align}
b &=\frac{\sigma_{8,g}}{\sigma_{8}},\textnormal{where}\\
\sigma_{8,g}^{2}&=\frac{72(r_{0}/8h^{-1}\textnormal{cMpc})^{\gamma}}{(3-\gamma)(4-\gamma)(6-\gamma)2^{\gamma}}
\end{align}

\section{Results}\label{section:results}

\subsection{Angular Correlation Function and Bias}

The right-most upper panel of Figure~\ref{figure:tpcf-6.5+} shows the combined ACF at $\overline{z}=7.2$. There is a clear clustering signal, detected at high confidence ($\gtrsim4\sigma$), and corresponding to a clustering length $r_{0}=\roseven h^{-1}$cMpc. For comparison, Figure~\ref{figure:tpcf-6.5+} also shows the ACF at $\overline{z}=3.9$, $\overline{z}=4.9$ and $\overline{z}=5.9$. Table \ref{table:values} summarizes our estimates of $A_{w}$, $r_{0}$ and $b$ for these ACF measurements. Figure~\ref{figure:bias} illustrates the evolution of the galaxy bias with redshift, and compares the bias of dark-matter halos at different masses computed with the \citet{sheth1999} formalism. The bias at $\overline{z}=7.2$ is $b=\biasseven$, with a clear increase compared to the lower redshift measurements. The evolution of the clustering strength with redshift broadly follows the increase expected at an approximately constant dark-matter halo mass of $M\sim10^{10.5}~$\msun out to $z\sim6$, followed by a possible increase to $M\sim10^{11}~$\msun at $z\sim7$ ($\lesssim2\sigma$ confidence). The near-constant trend between $z=5.9$ and $z=3.8$ is consistent with empirical models \citep{trenti2013, tacchella2013}, which predict no evolution or a mild decrease in halo masses at fixed UV-luminosity. In this respect, the $\overline{z}=7.2$ result is unexpected and could potentially indicate a change in star formation efficiency or clustering properties as the Universe becomes more neutral, but the uncertainty is still large.

To investigate luminosity dependence, we split galaxies in our samples into bright and faint subsamples. We cut at the median luminosity of the $\overline{z}=7.2$ sample, $M_{UV}=-19.4$. Since LF shape evolution approximately cancels out the evolution in the $k$-corrected distance modulus, this median cut is within $0.1$ mag of the median of each sample from $z=3.8$ to $z=7.8$ allowing us to explore bias variation at near-constant UV luminosity\footnote{Note that the cut is approximate at the level of $\lesssim0.1$ mag., since the distance modulus is computed at the mean redshift for each dropout sample}. At $\overline{z}=7.2$, the clustering strength in the bright sample is $r_{0}=\rosevenbright h^{-1}$cMpc, while the faint sample has $r_{0}=\rosevenfaint h^{-1}$cMpc, smaller, but consistent with no luminosity evolution within the measurement uncertainty. The presence of a luminosiy trend is more apparent at $\overline{z}=5.9$, where we observe galaxies in the bright subsample ($\overline{M}_{UV}=-20.2$) residing in more massive haloes ($M=10^{11.0^{+0.4}_{-0.6}}$\msun) than those in the faint sample ($\overline{M}_{UV}=-18.6$, $M=10^{9.0^{+0.9}_{-2.1}}$\msun). This is consistent with a constant mass-to-light ratio with luminosity in the $\overline{z}=5.9$ sample $\frac{\textnormal{dlogM}}{\textnormal{dlogL}}\sim1.2^{+1.8}_{-0.8}$. At lower redshift we observe a similar behavior (see also \citealt{lee2006}), with a high-confidence measurement at $\overline{z}=3.8$ (see Fig.~\ref{figure:bias} for the bias difference between LBGs in the bright and faint subsamples at $\overline{z}=3.8$ and $\overline{z}=5.9$). 

\subsection{Uncertainties and Analysis Validation}\label{sec:errors}

The uncertainties presented here derive from bootstrap resampling. These may underestimate the true uncertainty because they are derived at fixed $\beta$ and cosmic variance is not accounted for. To estimate cosmic variance, we use field-to-field variations in the bias, which are shown in Figure~\ref{figure:field_to_field}. Based on a $\chi^2$ analysis of the residuals between all-fields versus individual determination of the bias, we estimate that for each \emph{individual field} measurement in the combined sample (left panel), cosmic variance uncertainty is comparable to the random error (excluding $z=3.8$). The impact on the all-field determination is mitigated by analyzing independent regions in the sky, and if we conservatively assume only two independent pointings (GOODS-N and S), we get that the systematic cosmic variance uncertainty is $\Delta b=\pm1.0$ for our new measurement at $z=7.2$. The current high-$z$ subsamples have larger random errors, thus the impact of cosmic variance is negligible compared to the random errors. We also carried out our analysis assuming different values for $\beta$. At $z=7.2$, we obtain that $\Delta\beta=\pm0.1$ corresponds to $\Delta b\sim\mp0.9$. With a higher $\beta=0.8$, we would derive $b_{\overline{z}=7.2}=6.94^{+0.76}_{-0.85}$. We investigate the impact of changing the mean and width of $N(z)$ and find that shifting $N(z)$ by $\Delta z=\pm0.2$ changes our bias estimate by $\Delta b=\pm0.1$ at each redshift interval. A $10\%$ change to the width of $N(z)$ changes our bias estimate by $\Delta b=\pm0.2$.

To ensure robustness of our analysis against the combining of catalogs, we performed several checks. We first split the $\overline{z}=7.2$ sample into $z$-dropouts ($\overline{z}=6.8$) and $Y$-dropouts ($\overline{z}=7.9$), obtaining $A_{w}=\Awsixpointeight$, $r_{0}=\rosixpointeight h^{-1}$cMpc, $b=\biassixpointeight$ at $\overline{z}=6.8$, and $A_{w}=\Awsevenpointeight$, $r_{0}=\rosevenpointeight h^{-1}$cMpc, $b=\biassevenpointeight$ at $\overline{z}=7.8$. Additionally, we restrict the analysis to the B+14 sample, finding $A_{w}=\AwsevenBouwensOnly$, $r_{0}=\rosevenBouwensOnly h^{-1}$cMpc, $b=\biassevenBouwensOnly$. These determinations are all consistent with results from the full sample, demonstrating absence of systematic errors introduced by combining heterogeneous datasets.

To investigate the impact of multiple halo occupancy, which we are neglecting in our analysis but may affect the ACF in the innermost $\sim10''$ (see \citealt{lee2006}), we carried out our maximum likelihood determination of the ACF by excluding datapoints in the innermost $12.5''$. At $z=7.2$ we obtain $b=9.03^{+1.12}_{-1.29}$ for the full sample, which is completely consistent with $b=8.6\pm1.0$ from the full analysis. The lack of evidence for multiple halo occupation is expected, since multiple galaxies per halo are expected only when $M\gtrsim10^{12}~\mathrm{M_{\sun}}$ (very rare at $z>6.5$). 

Finally, we verified that our low-$z$ analysis is consistent with previous published investigations. The correlation lengths of $B$ and $V$ dropouts are in general agreement with those presented in \citet{lee2006}, who found $r_{0}=2.9^{+0.2}_{-0.2}$cMpc at $z\sim3.8$ and $r_{0}=4.4^{+0.5}_{-0.5}$cMpc at $z\sim4.9$, both with a fixed slope of $\beta=0.6$, for their largest sample at each redshift. Our results at $z\sim5.9$ are consistent with those presented in \citet{overzier2006}, who found $r_{0}=4.5^{+2.1}_{-3.2}$ in the GOODS fields.

\subsection{Duty Cycle}

The duty cycle of LBGs is defined as the fraction of dark-matter halos hosting UV-bright LBGs and is linked to the star formation efficiency realized in the halos. The duty cycle depends on the occupation efficiency of dark matter haloes and the timescale that galaxies remain visible at ultraviolet wavelengths. Our clustering measurements allow us to estimate duty cycles by comparing the observed bias with that predicted by ``abudance matching'', which connects luminosity functions to the dark-matter halo mass functions (e.g., \citealt{vale04}) by matching objects at the same comoving density. Abundance matching is carried out assuming only a fraction  $\epsilon_{DC}\leq1$ of halos hosts a UV-bright galaxy. The number density of haloes above a mass $M_{h}$ is matched to the number density of galaxies above a luminosity $L_{g}$ using the duty cycle, $\epsilon_{DC}\leq1$ \citep{trenti2010}:
\beq
\epsilon_{DC}\int^{+\infty}_{M_{h}}n(M_{h},z)dM_{h} = \int_{L_{g}}^{+\infty}\psi(L,z)dL
\eeq
where $\psi(L,z)$ is the luminosity function at redshift $z$, $n(M_{h},z)$ is the halo mass function as redshift $z$ and $\epsilon_{DC}$ is the duty cycle. We construct a mass-luminosity relation at $\overline{z}=7.2$ from abundance matching between the \citeauthor{sheth1999} mass function and a Schechter LF, defined as,
\beq
n(M)=\phi^{\star}\frac{\textrm{ln}(10)}{2.5}10^{-0.4(M-M_{\star})(\alpha+1)}e^{10^{-0.4(M-M^{\star})}}
\eeq
with $\phi^{\star}=0.82\times10^{-3}$, $M^{\star}=-20.2$, $\alpha=-1.86$ \citep{bouwens2011}. Matching luminosity with dark-matter halo mass gives us the average host halo mass of the galaxies, and thus a halo bias from the \citeauthor{sheth1999} formalism which depends on $\epsilon_{DC}$. This is shown in the bottom panels of Figure~\ref{figure:tpcf-6.5+}. By taking the intersection between the bias inferred from abudance matching and from clustering analysis, we derive the value of $\epsilon_{DC}$.

From Figure~\ref{figure:tpcf-6.5+}, we see that at $\overline{z}=7.2$ duty cycles near $\epsilon_{\textnormal{DC}}=1.0$ are favored for the total sample, but all duty cycle values from abundance matching are in mild tension with the observed bias (which is too high). This discrepancy could be due to the systematic errors present in the analysis (cosmic variance and/or fixed $\beta$). As discussed in Section \ref{sec:errors}, the tension at $\overline{z}=7.2$ is resolved by using a value of $\beta=0.8$. Including the fixed-$\beta$ and cosmic variance uncertainties will approximately double the total uncertainty, compared to the random error $1\sigma$ interval shown in Figure \ref{figure:bias}.

At $z\sim7$ our duty cycle measurement contrasts with the theoretical expectation from \citet{wyithe2013} who noted that for $z>4$ a low duty cycle ($\epsilon_{DC}\sim0.1-0.15$) would be required to resolve the tension between the observed star-formation rate with the comparatively low observed total stellar mass. This prediction is consistent with our measurements at $4\lesssim z\lesssim6$. Another explanation for the discrepancy may be the effect of reionization on galaxy bias measurements. \citet{wyithe2007} showed that reionization, which occurs earlier in overdense regions, can enhance the observed galaxy bias in flux-limited samples. Early reionization of overdense regions leads to different star-formation histories due to heating of the IGM, and hence a different luminosity at fixed stellar mass. This results in overdense regions hosting a younger, brighter population of LBGs, which causes the observed galaxy bias to be overestimated. At $z\sim7$, this effect may contribute at the $\sim10\%$ level ($\Delta b\sim1$) \citep{wyithe2007}, which is similar to the discrepancy observed in the lower panels of Figure \ref{figure:tpcf-6.5+}. This effect also results in an overestimation of the halo mass from the bias (a factor of $\sim3$ for halos with $M\sim10^{11}$\msun). 

Previous studies, such as \citet{lee2009}, found lower duty cycles at lower redshifts. As starburts remain UV-bright for $\sim100$Myr, a shorter halo assembly time at high redshifts results in a higher duty cycle, as more haloes will be seen to host UV-bright galaxies (e.g. see the top-right panel of Fig.~1 in \citealt{trenti2010}). This may qualitatively explain the increase in duty cycle observed with redshift. However, the current uncertainties in the analysis make it difficult to draw any firm conclusion on the redshift evolution of $\epsilon_{\textnormal{DC}}$.

\section{Conclusion}
\label{section:conclusion}
In this paper we carry out the first clustering analysis of LBG galaxies at $z\geq6.5$, taking advantage of a combined sample of $N=743$ candidates from the XDF, GOODS-S and GOODS-N legacy fields. We detect a positive clustering signal with $\gtrsim4\sigma$ confidence, finding the real-space correlation length and bias of our $\overline{z}=7.2$ sample to be $r_{0}=\roseven h^{-1}$cMpc and $b=\biasseven$ respectively. This result, and its interpretation, provides fundamental insight into several aspects of galaxy formation and evolution during the epoch of reionization:
\begin{itemize}
\item We show that the bias of galaxies at fixed UV-luminosity clearly increases with redshift, from $b\sim3.0$ at $z\sim4$ to $b\sim8.6$ at $z\gtrsim7$, implying that LBGs of a fixed luminosity reside in dark-matter halos with similar masses at $4\lesssim z\lesssim6.5$, followed by a moderate halo mass increase at $z\gtrsim7$. While the current uncertainty on the bias is still too large to provide stringent tests of models of galaxy formation, the tension between predicted and measured bias at $z\gtrsim7$ could be linked to the effect of reionization \citep{wyithe2007}. Overall our analysis clearly demonstrates the potential of clustering analysis in future datasets; 

\item We constrain the mass of dark-matter halos hosting the LBGs observed by Hubble, finding that $z\gtrsim7$ galaxies live in halos with $M\sim10^{11}$~\msun. Since dark-matter halos grow by around two orders of magnitude in mass from $z\sim7$ to $z\sim0$ based on Extended Press-Schechter modeling (e.g., Fig. 7 in  \citealt{trenti12a}), this implies that the observed population of LBGs at $z>4$ will predominantly end up in a group or small-cluster environment by the present time;


\item We observe luminosity-dependent clustering within the $\overline{z}=5.9$ sample when split at $M_{UV}=-19.4$. The change in halo bias between these two samples implies a mass-to-light ratio of $\frac{\textnormal{dlogM}}{\textnormal{dlogL}}\sim1.2^{+1.8}_{-0.8}$.  A similar trend is present for galaxies at $z\geq6.5$, but with larger uncertainty;

\item Finally, we provide a first constraint on the duty cycle of $z\geq6.5$ LBGs, finding values close to unity (Figure~\ref{figure:tpcf-6.5+}) and a possible evolution of the duty cycle with redshift. Such evolution is likely related to the shorter halo assembly times at increasing redshift;

\end{itemize}

These results highlight the significance of the progress in the study of galaxies during the first Gyr, made possible by the installation of WFC3 on Hubble and by the telescope time devoted to deep surveys such as XDF and CANDELS. The measure of galaxy clustering has been a fundamental tool in galaxy formation and evolution at lower redshift starting from the establishment of the cosmic web and its connection to dark-matter halos (e.g. \citealt{davis85}). Now, we have demonstrated how this measurement is possible for samples of objects during reionization: While the uncertainty on the galaxy bias at $z\geq6.5$ is still significant, upcoming programs such as the Frontier Fields Initiative will provide even stronger constraints on the connection between baryons and dark-matter in the youth of the Universe. 

\smallskip
\smallskip

\acknowledgements

We thank the anonymous referee for their useful suggestions and comments that have improved the manuscript. Support for this work was partially provided by the Australian Research Council through an Australian Laureate Fellowship FL110100072 and through Discovery Project DP140103498; by the European Commission through the Marie Curie Career Integration Fellowship PCIG12-GA-2012-333749; and by NASA through Hubble Fellowship grant HF-51278.01 and grants HSTGO-12905, HSTGO-12572, and HSTGO-11563.

\begin{figure*}
\begin{center}
\includegraphics[scale=0.21]{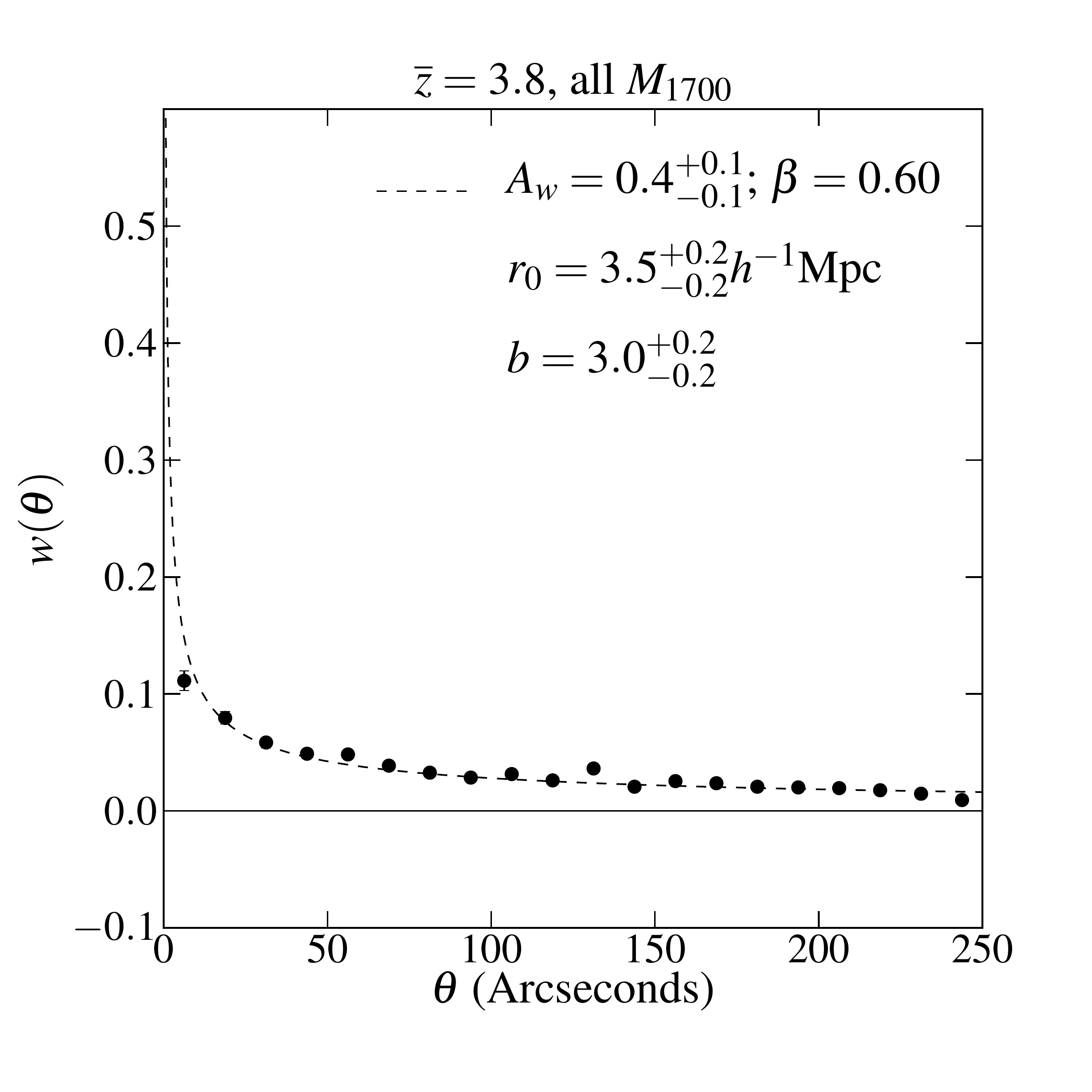}
\includegraphics[scale=0.21]{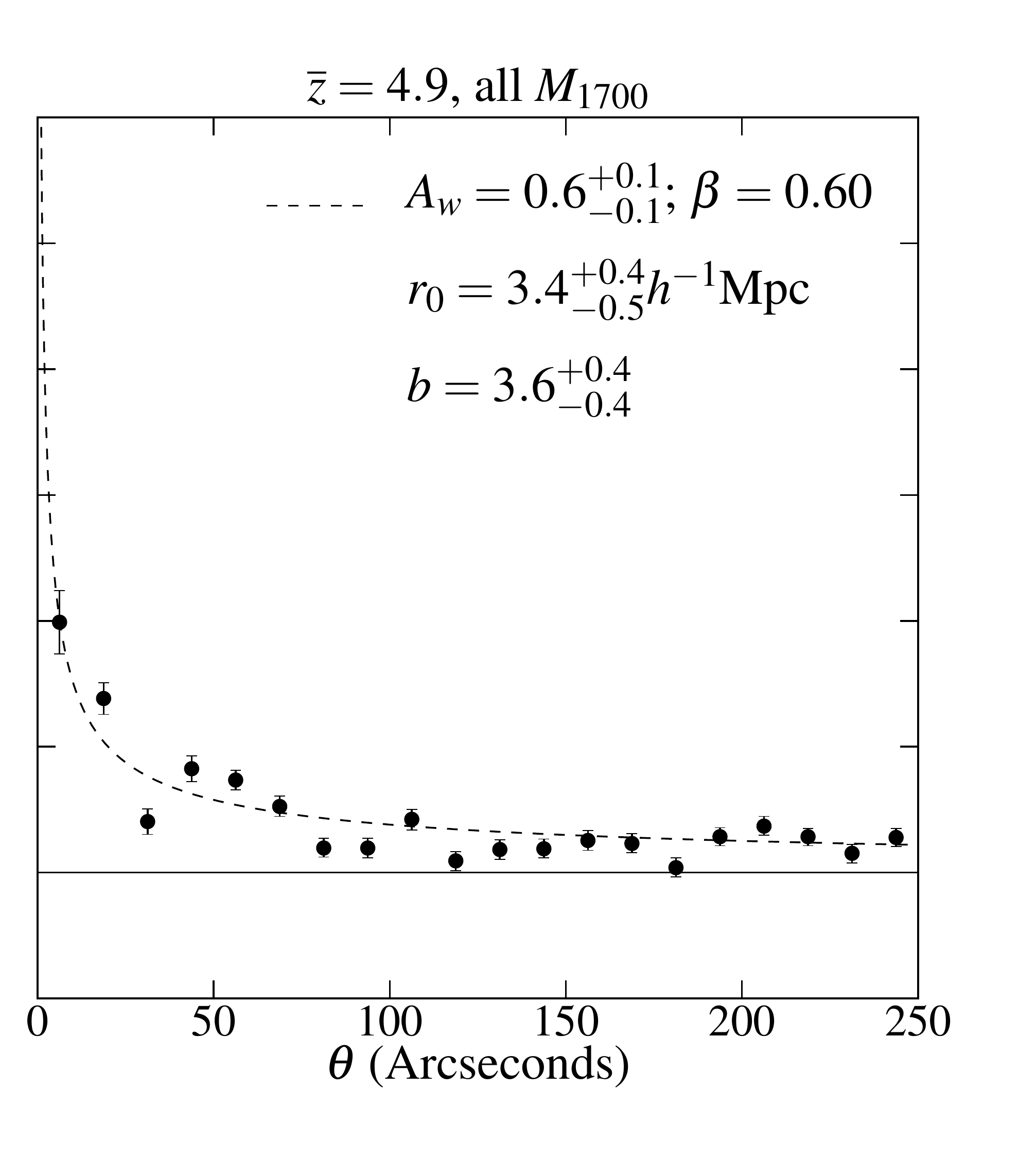}
\includegraphics[scale=0.21]{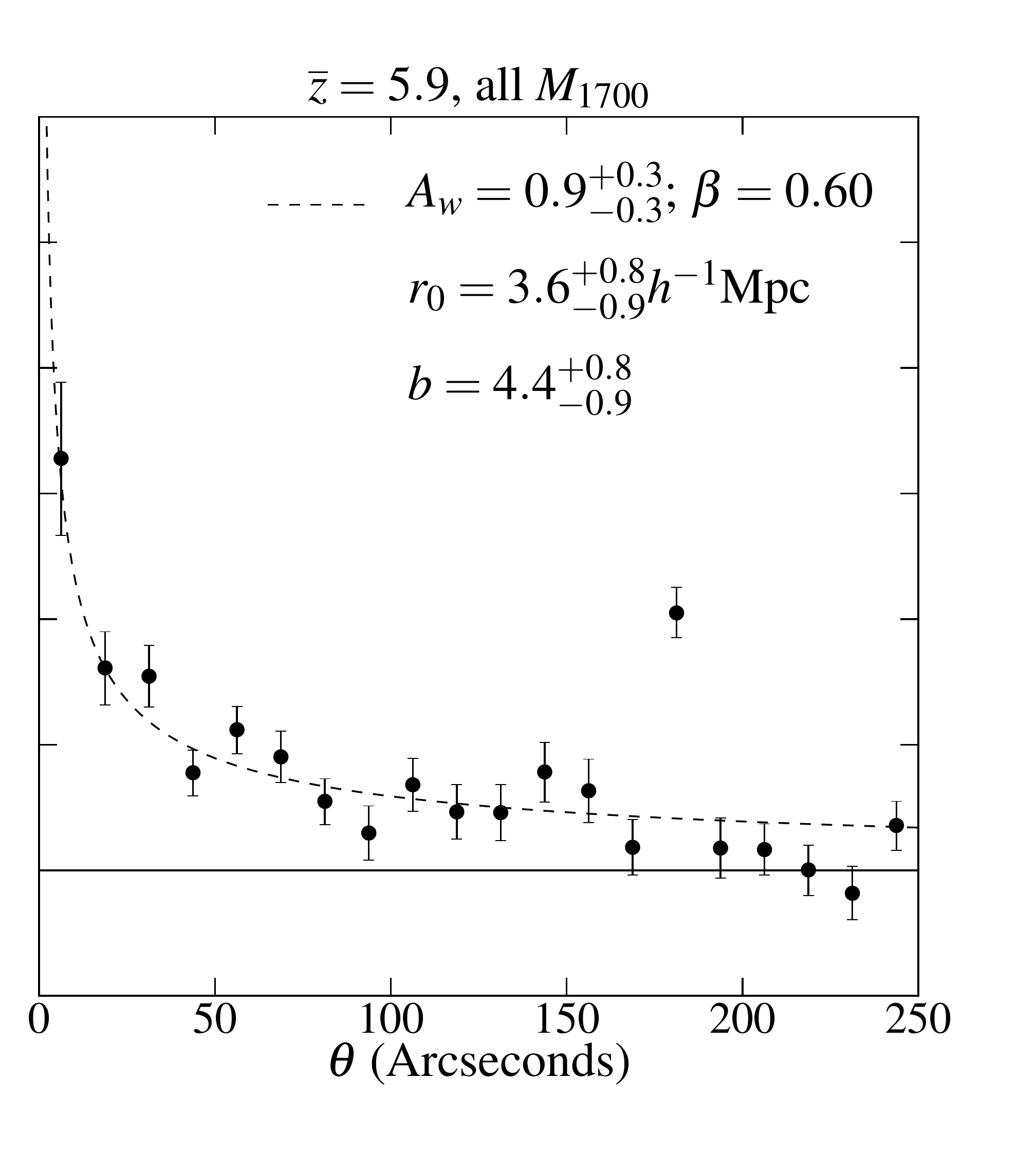}
\includegraphics[scale=0.21]{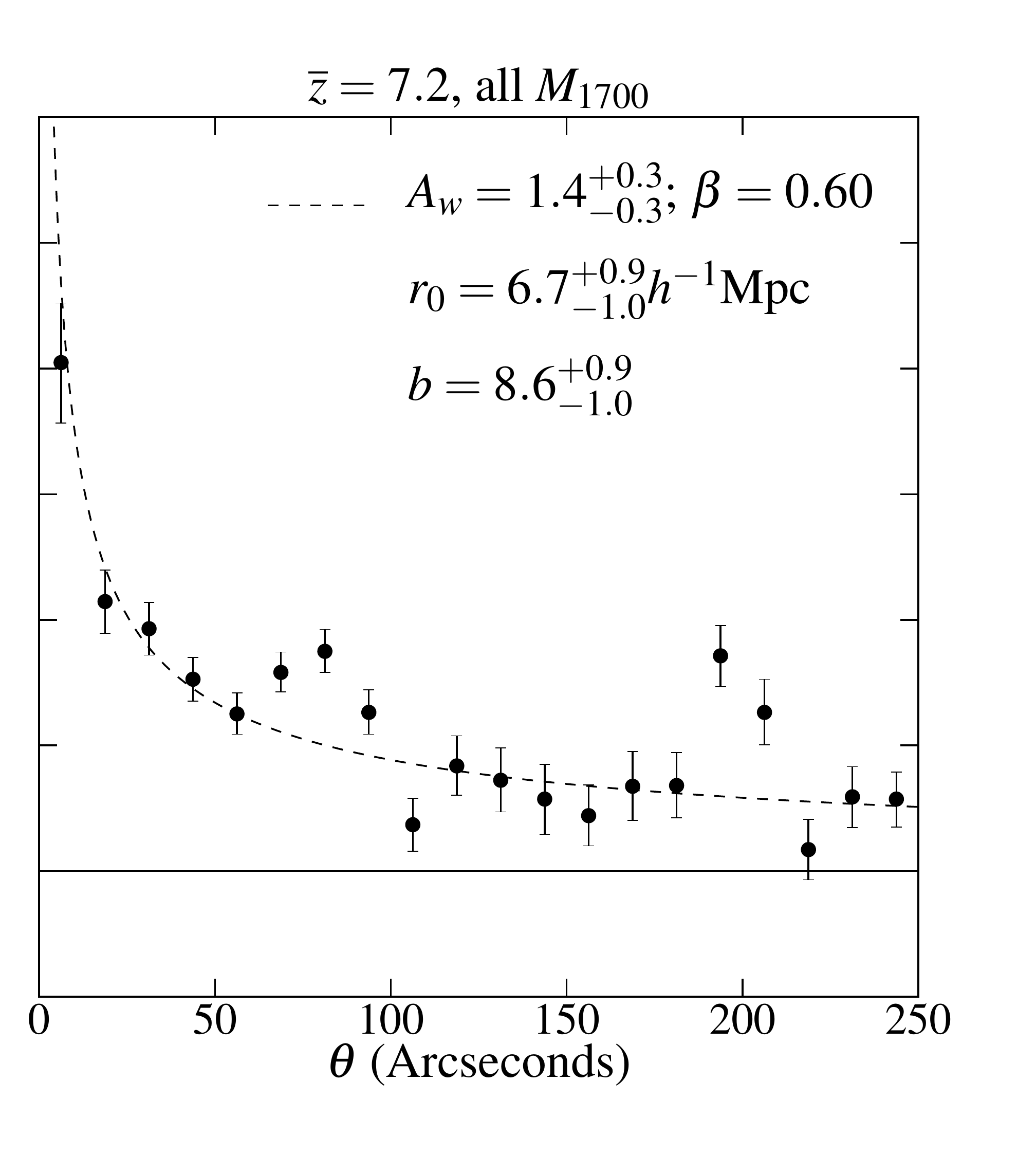}\\
\includegraphics[scale=0.2125]{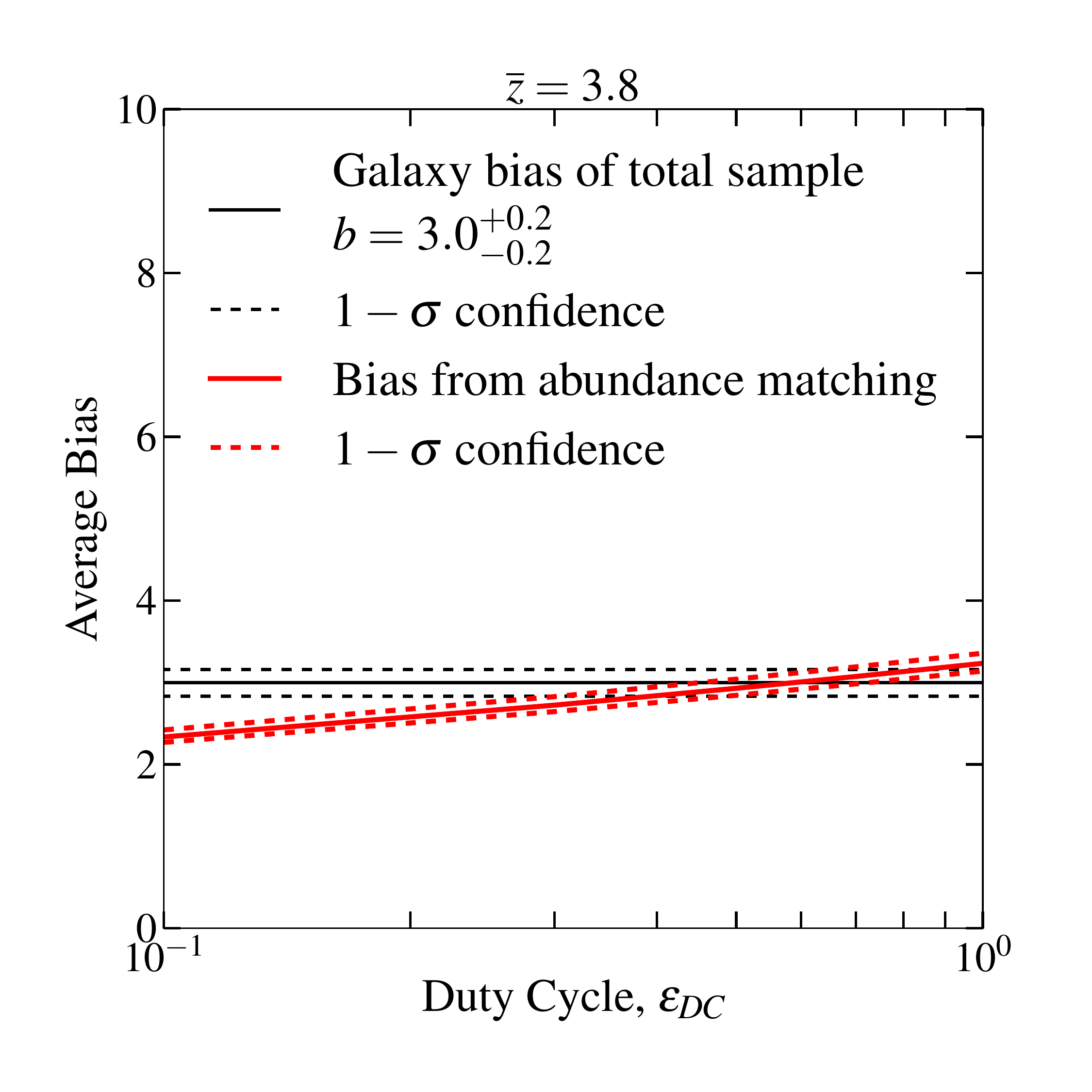}
\includegraphics[scale=0.2125]{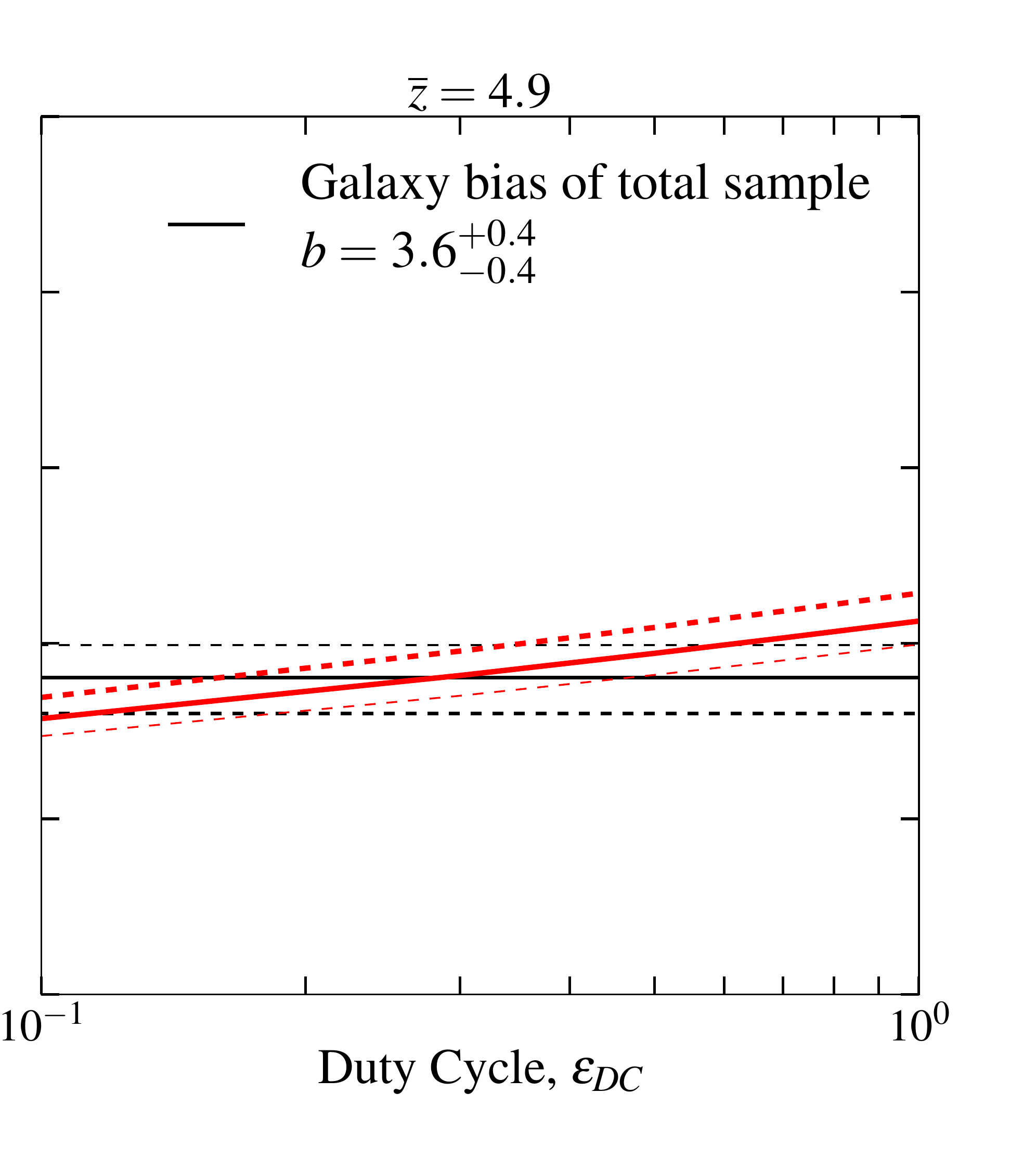}
\includegraphics[scale=0.2125]{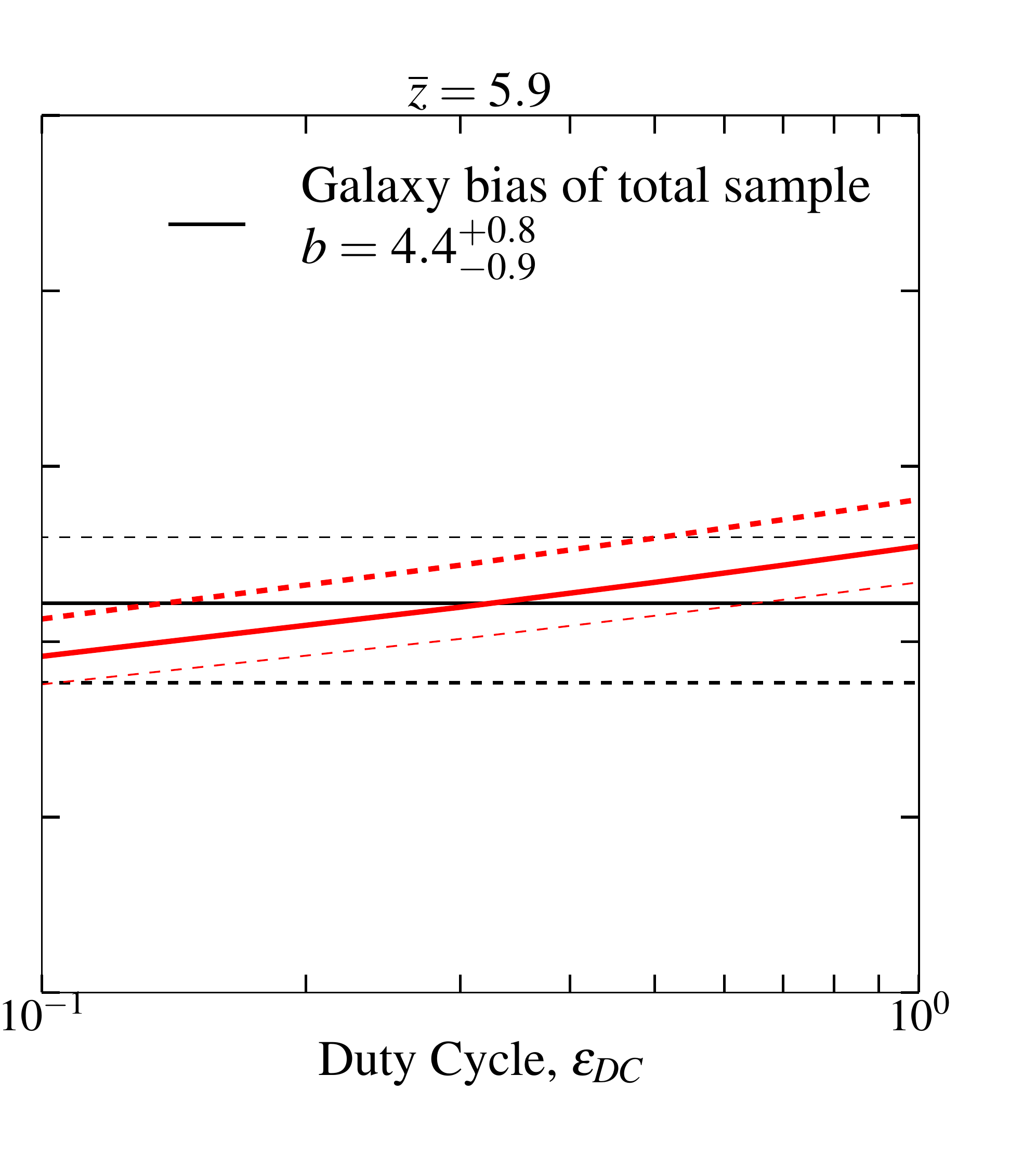}
\includegraphics[scale=0.2125]{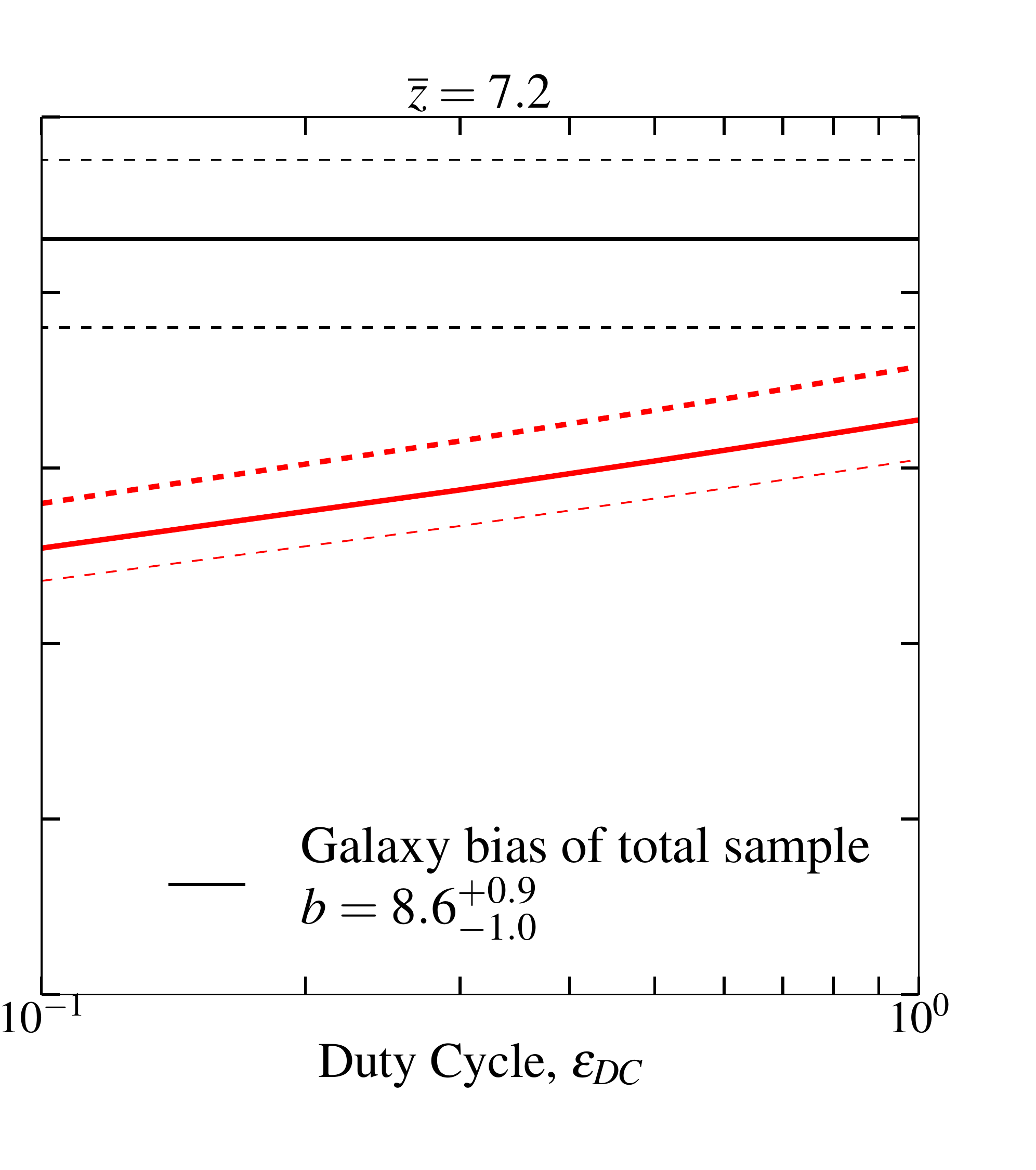}
\caption{\textit{Top:} ACF for LBGs at $\overline{z}=3.8$,  $\overline{z}=4.9$, $\overline{z}=5.9$ and $\overline{z}=7.2$ (left to right). We detect the first clustering signals at $z\sim7$ ($\gtrsim4\sigma$ confidence). \textit{Bottom:} Average galaxy bias in the total sample at $\overline{z}=3.8$,  $\overline{z}=4.9$, $\overline{z}=5.9$ and $\overline{z}=7.2$ (left to right) computed using abundance matching from the \citet{sheth1999} mass function as a function of duty cycle (red). The $1$-$\sigma$ uncertainty of the bias from abundance matching is estimated from the LF Schechter-fit uncertainty.}
\label{figure:tpcf-6.5+}
\end{center}
\end{figure*}
%
%
\begin{figure}
\begin{center}
\includegraphics[scale=0.4]{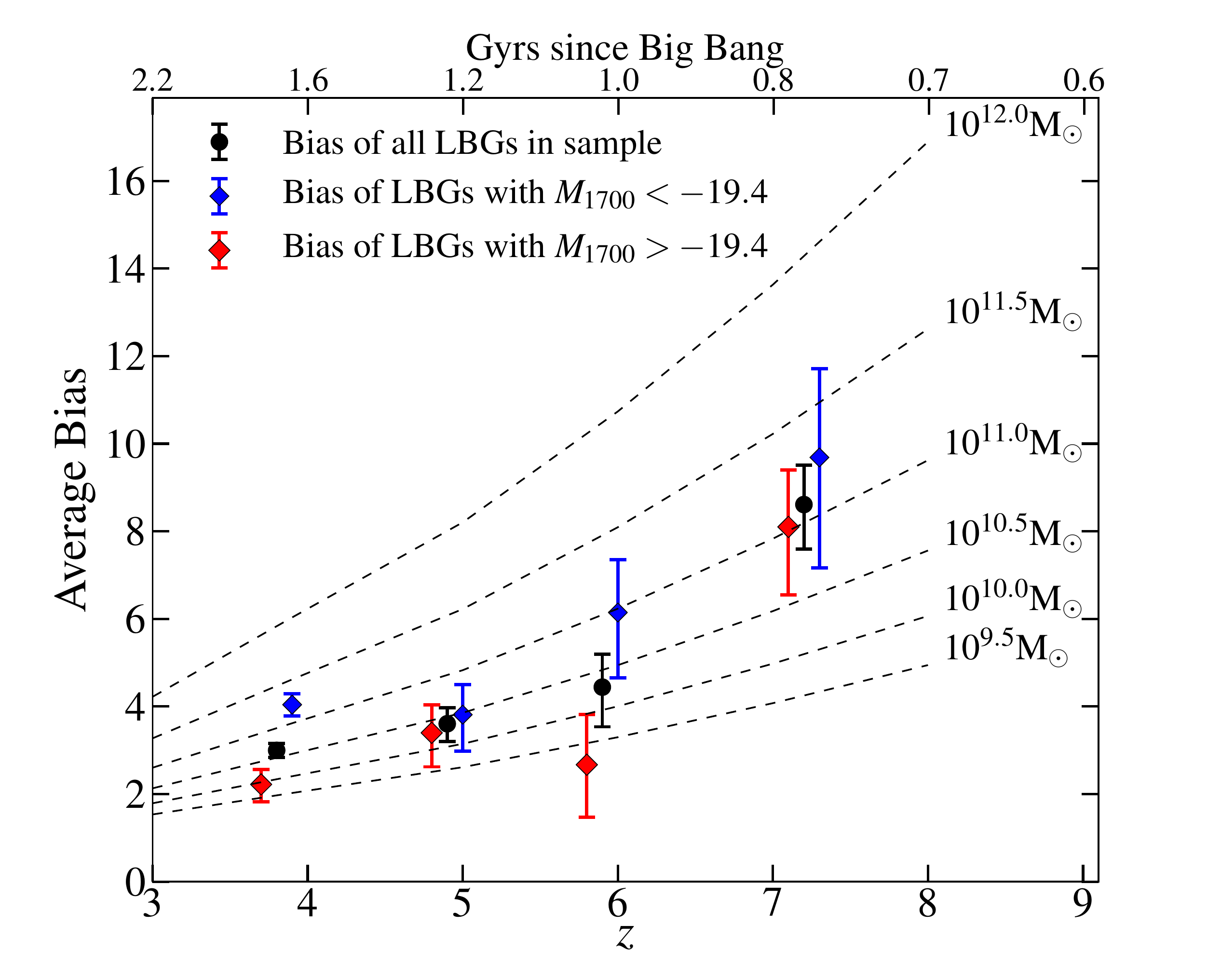}
\caption{The bias evolution as a function of redshift from $3.8<z<7.2$ for our total (black circles), bright (blue diamonds, offset by $+0.1$ in $z$) and faint (red diamonds, offset by $-0.1$ in $z$) samples plotted against the dark-matter halo bias from the \citet{sheth1999} mass function.} 
\label{figure:bias}
\end{center}
\end{figure}
%

\begin{figure}
\begin{center}
\includegraphics[scale=0.36]{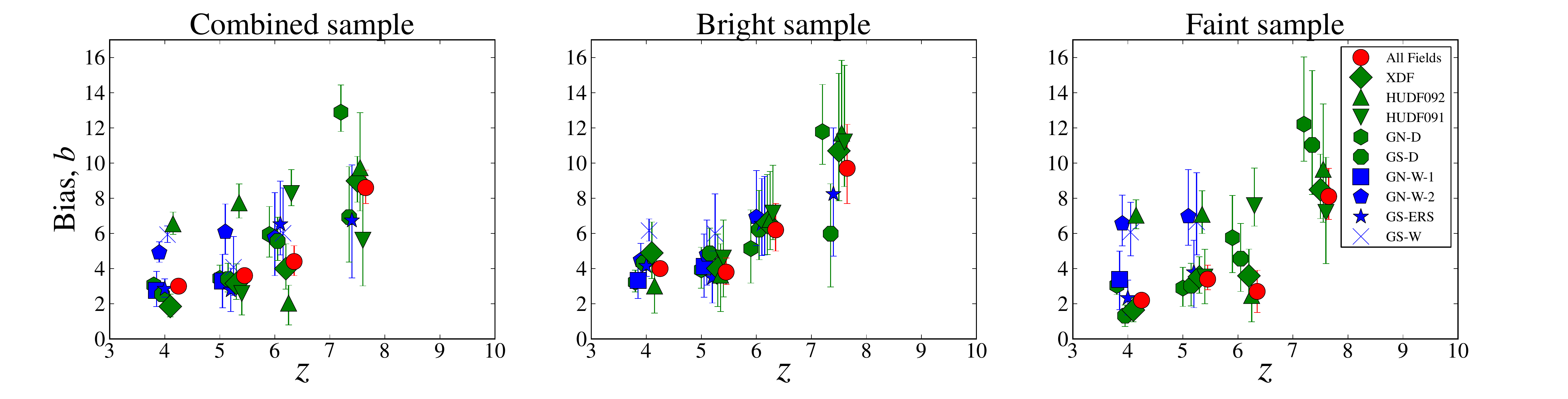}
\caption{Bias determination in individuals fields for the full (left), bright (center), and faint (right) samples shown as blue points with error-bars (slightly shifted in redshift for clarity), and compared to the combined maximum likelihood fit (red points).}
\label{figure:field_to_field}
\end{center}
\end{figure}


\begin{thebibliography}{26}
\expandafter\ifx\csname natexlab\endcsname\relax\def\natexlab#1{#1}\fi

\bibitem[{Adelberger {et~al}\mbox{.}(1998)Adelberger, Steidel, Giavalisco,
  Dickinson, Pettini, \& Kellogg}]{adelberger1998}
Adelberger K.~L., Steidel C.~C., Giavalisco M., Dickinson M., Pettini M.,
  Kellogg M., 1998, The Astrophysical Journal, 505, 18

\bibitem[{Adelberger {et~al}\mbox{.}(2005)Adelberger, Steidel, Pettini,
  Shapley, Reddy, \& Erb}]{adelberger2005}
Adelberger K.~L., Steidel C.~C., Pettini M., Shapley A.~E., Reddy N.~A., Erb
  D.~K., 2005, The Astrophysical Journal, 619, 697

 \bibitem[{Arnouts {et~al}\mbox{.}(1999)Arnouts, Cristiani, Moscardini,
  Matarrese, Lucchin, Fontana, \& Giallongo}]{arnouts1999}
Arnouts S., Cristiani S., Moscardini L., Matarrese S., Lucchin F., Fontana A.,
  Giallongo E., 1999, Monthly Notices of the Royal Astronomical Society, 310,
  540

\bibitem[Berlind \& Weinberg(2002)]{berlind2002} Berlind, A.~A. and Weinberg, D.~H. 2002, \apj, 575, 587

\bibitem[{Bertin \& Arnouts(1996)}]{bertin1996}
Bertin E., Arnouts S., 1996, Astronomy and Astrophysics Supplement Series, 117,
  393

\bibitem[{Bouwens {et~al}\mbox{.}(2011)Bouwens, Illingworth, Oesch, Labb{\'e},
  Trenti, van Dokkum, Franx, Stiavelli, Carollo, Magee,
  {et~al.}}]{bouwens2011}
Bouwens R. {et~al.}, 2011, The Astrophysical Journal, 737, 90

\bibitem[Bouwens et al.(2014)]{bouwens2014} Bouwens R. et al. 2014, arXiv1403.4295B [B+14]

\bibitem[{Bradley {et~al}\mbox{.}(2012)Bradley, Trenti, Oesch, Stiavelli, Treu,
  Bouwens, Shull, Holwerda, \& Pirzkal}]{bradley2012}
Bradley L. {et~al.}, 2012, The Astrophysical Journal, 760, 108

\bibitem[Davis et al.(1985)]{davis85} Davis M., Efstathiou G., Frenk C. S., \& White S. D. M. 1985, The Astrophysical Journal, 292, 37

\bibitem[{Planck Collaboration(2013)Collaboration, Ade, Aghanim,
  Armitage-Caplan, Arnaud, Ashdown, Atrio-Barandela, Aumont, Baccigalupi,
  Banday, {et~al.}}]{collaboration2013}
Collaboration P. {et~al.}, 2013, arXiv preprint arXiv:1303.5076

\bibitem[{Giavalisco \& Dickinson(2001)}]{giavalisco2001}
Giavalisco M., Dickinson M., 2001, The Astrophysical Journal, 550, 177

\bibitem[Grogin et al.(2011)]{grogin2011} Grogin, N. et al. (2011), The Astrophysical Journal Supplement, 197, 35

\bibitem[{Illingworth {et~al}\mbox{.}(2013)Illingworth, Magee, Oesch, Bouwens,
  Labbe, Stiavelli, van Dokkum, Franx, Trenti, Carollo,
  {et~al.}}]{illingworth2013}
Illingworth G. {et~al.}, 2013, The Astrophysical Journal Supplement, 209, 6

\bibitem[{Jaacks {et~al}\mbox{.}(2012)Jaacks, Choi, Nagamine, Thompson, \&
  Varghese}]{jaacks2012}
Jaacks J., Choi J.-H., Nagamine K., Thompson R., Varghese S., 2012, Monthly
  Notices of the Royal Astronomical Society, 420, 1606

\bibitem[{Lacey {et~al}\mbox{.}(2011)Lacey, Baugh, Frenk, \&
  Benson}]{lacey2011}
Lacey C., Baugh C., Frenk C., Benson A., 2011, Monthly Notices of the Royal
  Astronomical Society, 412, 1828

\bibitem[{Landy \& Szalay(1993)}]{landy1993}
Landy S.~D., Szalay A.~S., 1993, The Astrophysical Journal, 412, 64

\bibitem[{Lee {et~al}\mbox{.}(2006)Lee, Giavalisco, Gnedin, Somerville,
  Ferguson, Dickinson, \& Ouchi}]{lee2006}
Lee K.-S., Giavalisco M., Gnedin O.~Y., Somerville R.~S., Ferguson H.~C.,
  Dickinson M., Ouchi M., 2006, The Astrophysical Journal, 642, 63
 
\bibitem[{Lee {et~al}\mbox{.}(2009)Lee, Giavalisco, Conroy, Wechsler, Ferguson, 
Somerville, Dickinson, \& Urry}]{lee2009}
Lee K.-S., Giavalisco M., Conroy C., Wechsler R.~H., Ferguson H.~C., 
Somerville R.~S., Dickinson M.~E., Urry C.~M, 2009, The Astrophysical Journal, 695, 368

\bibitem[{Ling {et~al}\mbox{.}(1986)Ling, Barrow, \&
  Frenk}]{ling1986}
Ling E.~N., Barrow J.~D., Frenk C., 1986, Monthly Notices of the Royal
  Astronomical Society, 223, 21P

\bibitem[{McLure {et~al}\mbox{.}(2013)McLure, Dunlop, Bowler, Curtis-Lake,
  Schenker, Ellis, Robertson, Koekemoer, Rogers, Ono, {et~al.}}]{mclure2012}
McLure R. {et~al.}, 2013, Monthly Notices of the Royal Astronomical Society,
  432, 2696

\bibitem[{Oesch {et~al}\mbox{.}(2012)Oesch, Bouwens, Illingworth, Gonzalez,
  Trenti, van Dokkum, Franx, Labb{\'e}, Carollo, \& Magee}]{oesch2012}
Oesch P. {et~al.}, 2012, The Astrophysical Journal, 759, 135

\bibitem[{Oke(1974)}]{oke1974}
Oke J., 1974, The Astrophysical Journal Supplement Series, 27, 21

\bibitem[{Ouchi {et~al}\mbox{.}(2004)Ouchi, Shimasaku, Okamura, Furusawa,
  Kashikawa, Ota, Doi, Hamabe, Kimura, Komiyama, {et~al.}}]{ouchi2004}
Ouchi M. {et~al.}, 2004, The Astrophysical Journal, 611, 685

\bibitem[{Overzier {et~al}\mbox{.}(2006)Overzier, Bouwens, Illingworth, \&
  Franx}]{overzier2006}
Overzier R.~A., Bouwens R.~J., Illingworth G.~D., Franx M., 2006, The
  Astrophysical Journal Letters, 648, L5

\bibitem[{Peebles(1980)}]{peebles1980}
Peebles P. J.~E., 1980, The large-scale structure of the universe. Princeton
  university press

\bibitem[{Sheth \& Tormen(1999)}]{sheth1999}
Sheth R.~K., Tormen G., 1999, Monthly Notices of the Royal Astronomical
  Society, 308, 119
  
 \bibitem[{Steidel {et~al}\mbox{.}(1996)Steidel, Giavalisco, Dickinson, \&
  Adelberger}]{steidel1996}
Steidel C.~C., Giavalisco M., Dickinson M., Adelberger K.~L., 1996, Arxiv
  preprint astro-ph/9604140

\bibitem[{Tacchella {et~al}\mbox{.}(2013)Tacchella, Trenti, \&
  Carollo}]{tacchella2013}
Tacchella S., Trenti M., Carollo C.~M., 2013, The Astrophysical Journal
  Letters, 768, L37

\bibitem[{Trenti {et~al}\mbox{.}(2010)Trenti, Stiavelli, Bouwens, Oesch, Shull,
  Illingworth, Bradley, \& Carollo}]{trenti2010}
Trenti M., Stiavelli M., Bouwens R., Oesch P., Shull J., Illingworth G.,
  Bradley L., Carollo C., 2010, The Astrophysical Journal Letters, 714, L202

\bibitem[{Trenti {et~al}\mbox{.}(2011)Trenti, Bradley, Stiavelli, Oesch, Treu,
  Bouwens, Shull, MacKenty, Carollo, \& Illingworth}]{trenti2011}
Trenti M. {et~al.}, 2011, The Astrophysical Journal Letters, 727, L39

\bibitem[Trenti et al.(2012a)]{trenti12a} Trenti, M. et al. 2012,  The Astrophysical Journal Letters, 746, 55

\bibitem[{Trenti {et~al}\mbox{.}(2013)Trenti, Perna, \&
  Tacchella}]{trenti2013}
Trenti M., Perna R., Tacchella S., 2013, The Astrophysical Journal Letters,
  773, L22

\bibitem[Trenti \& Stiavelli(2008)]{trenti08} Trenti, M. and Stiavelli, M. 2008, The Astrophysical Journal, 676, 767

\bibitem[Vale \& Ostriker(2004)]{vale04} Vale, A. and Ostriker, J.~P. 2004, \mnras, 353, 189

\bibitem[{Windhorst {et~al}\mbox{.}(2011)Windhorst, Cohen, Hathi, McCarthy,
  Ryan~Jr, Yan, Baldry, Driver, Frogel, Hill, {et~al.}}]{windhorst2011}
Windhorst R.~A. {et~al.}, 2011, The Astrophysical Journal Supplement Series,
  193, 27
  
\bibitem[{Wyithe \& Loeb(2007)}]{wyithe2007}
Wyithe J. S.~B., Loeb A., 2007, Monthly Notices of the Royal Astronomical
  Society, 382, 921

\bibitem[{Wyithe \& Loeb(2013)}]{wyithe2013suppressed}
Wyithe J. S.~B., Loeb A., 2013, Monthly Notices of the Royal Astronomical
  Society, 428, 2741

\bibitem[{Wyithe {et~al}\mbox{.}(2013)Wyithe, Loeb, \&
  Oesch}]{wyithe2013}
Wyithe S., Loeb A., Oesch P., 2013, arXiv preprint arXiv:1308.2030

\end{thebibliography}
\end{document}